\documentclass[%
 reprint,
 amsmath,amssymb,
 aps,
]{revtex4-1}

\usepackage{graphicx}
\usepackage{caption}
\usepackage{subcaption}
\usepackage{dcolumn}
\usepackage{bm}

\begin{document}

\preprint{APS/123-QED}

\title{Role of zero-point effects in stabilizing the ground state structure of bulk Fe$_2$P}
\author{Soumya S. Bhat, Kapil Gupta, Satadeep Bhattacharjee and Seung Cheol Lee}
\email{seungcheol.lee@ikst.res.in}
\affiliation{Indo-Korea Science and Technology Center (IKST), Bangalore 560065, India}

\date{\today}
             
\begin{abstract}
Structural stability of Fe$_2$P is investigated in detail using first-principles calculations based on density functional theory. While the orthorhombic C23 phase is found to be energetically more stable, the experiments suggest it to be hexagonal C22 phase. In the present study, we show that in order to obtain the correct ground state structure of Fe$_2$P from the first-principles based methods it is utmost necessary to consider the zero-point effects such as zero-point vibrations and spin fluctuations. This study demonstrates an exceptional case where a bulk material is stabilized by quantum effects, which are usually important in low-dimensional materials. Our results also indicate the possibility of magnetic field induced structural quantum phase transition in Fe$_2$P, which should form the basis for further theoretical and experimental efforts.
 
\end{abstract}

\pacs{}
\maketitle
\section{\label{sec:level1}Introduction}
Fe$_2$P has attracted tremendous technological importance as it shows unique magnetic properties such as large uniaxial magnetocrystalline anisotropy {\cite{caron2013magnetocrystalline}}, magnetocaloric {\cite{geng2016microstructure,fries2017microstructural}} and magnetoelastic {\cite{gercsi2013magnetoelastic}} properties, due to which it finds wide use in industrial applications. In recent years, it has been also identified as potential electrocatalyst for hydrogen evolution reaction {\cite{ai2000oxidation,zhao2011hydrodeoxygenation}}. Besides industrial significance, Fe$_2$P is also interesting to geophysicists as a major component of meteorites in the form of barringerite and allabogdanite minerals {\cite{britvin2002allabogdanite}}.

In view of its importance, many theoretical and experimental efforts are devoted to study the magnetic and catalytic properties of Fe$_2$P {\cite{geng2016microstructure,fries2017microstructural,brock2008recent,costa2012large}}, yet its elemental property such as thermodynamic structural stability is hardly discussed in literature. At ambient conditions, Fe$_2$P crystallizes in hexagonal C22-type structure with space group $P\bar{6}2m$ {\cite{carlsson1973determination}}. An early high pressure and high temperature X-ray diffraction experiment on Fe$_2$P by Senateur {\it et al.} {\cite{senateur1976etude}}, disclosed the presence of orthorhombic C23-type polymorph of Fe$_2$P with space group {\it Pnma}, under compression. This was later confirmed by recent experiments {\cite{dera2008high,gu2013probing}}, where Fe$_2$P is reported to undergo structural phase transition from C22 to C23 structure under pressure of 8 GPa upon heating to 1400 K. On the contrary, first-principles calculations based on density functional theory (DFT) predicted C23 structure as the stable phase of Fe$_2$P {\cite{wu2010first}}. The discrepancy in the experimental observation and theoretical prediction of ground state structure of Fe$_2$P makes it essential to carry out a systematic analysis.

As stated earlier, in contrast to the experimental observation, theory predicts C23 structure as ground state of Fe$_2$P and there are no systematic studies on the structural stability of Fe$_2$P reported in literature. To fill this gap between the theory and experiment, we investigate the structural stability of Fe$_2$P using first-principles calculations based on DFT. Through comparison of structural parameters and magnetic properties of C22 and C23 phases, we analyze their consequences to stability. By determining full phonon spectra of C22 and C23 structures, we provide insights into the vibrational free energy contribution to the stability of C22 phase. Further, we determine ferromagnetic susceptibilities and compare zero-point spin fluctuation energies for both phases with implication to its ground state stability.
\section{Computational Details}
We performed first-principles calculations using the projector-augmented wave (PAW) {\cite{blochl1994projector}} method in the framework of DFT as implemented in VASP code {\cite{kresse1993ab,kresse1996efficient}}. The exchange-correlation energy of electrons is treated within a generalized gradient approximated functional (GGA) of the Perdew-Burke-Ernzerhof (PBE) {\cite{perdew1996generalized} parametrized form. Interactions between ionic cores and valence electrons are represented using PAW pseudopotentials, where 4{\it{s}}, 3{\it{d}} electrons for Fe and 3{\it s}, 3{\it p} electrons for P are treated as valence. Plane-wave basis set with kinetic energy cutoff of 600 eV and an energy convergence criteria of $10^{-6}$ eV are used. Uniform meshes of 9x9x15 and 8x16x8 k-points are used for Brillouin zone sampling of hexagonal and orthorhombic unit cells, respectively. Phonon frequencies across the Brillouin zone are obtained using finite-displacement approach as implemented in the Phonopy code {\cite{togo2008first}. We used 2x2x3 and 2x3x2 supercells for C22 and C23 structures, respectively. The forces induced by small displacements are calculated using VASP, with $10^{-8}$ eV as energy convergence criterion.
\section{Results and Discussion}
\subsection{Electronic structure}
For our calculations, initial structures are taken from experimental data {\cite{carlsson1973determination,senateur1976etude} and are optimized by full relaxation of the unit cell and atomic positions. The unit cell of hexagonal C22 phase composed of three formula units, with six Fe and three P atoms. The Fe atoms occupy two non-equivalent threefold symmetry sites Fe(I) and Fe(II), while the P atoms are situated on one twofold, P(I) and on one single-fold P(II) position in the crystal lattice. Fe(I) atom is surrounded by four P atoms with tetrahedral symmetry, whereas Fe(II) atom is surrounded by five P atoms, with pyramidal symmetry. The structure can be expressed as Fe3 triangles in the ab plane, with P occupying the alternate layers. The unit cell of orthorhombic C23 structure has four formula units and shares a similar pseudo-hexagonal arrangement of iron atoms as in C22 structure, but the arrangement of P atoms is different from that of the hexagonal structure in that half of the P(II) atoms are neighboring empty Fe3 triangles along the c-direction {\cite{gu2013probing}}.

Calculated ground state structural parameters and magnetic properties of Fe$_2$P for both C22 and C23 phases are listed in Table \ref{table1}, along with a comparison to earlier experimental data. For C22 structure, lattice parameters {\it a} and {\it c} are underestimated by $\sim1\%$. For C23 the error in the estimation of {\it a} and {\it c} is $\sim3\%$ and 2$\%$ respectively compared with the available experimental result \cite{senateur1976etude}, but the lattice constants are in good agreement with previous theoretical results {\cite{wu2010first}}. Further, computed lattice constants under hydrostatic pressure of ~30 GPa ({\it a} = 5.39, {\it b} = 3.45 and {\it c} = 6.31 \r{A}) agree well with the reported experimental values {\cite{dera2008high}} ({\it a} = 5.38, {\it b} = 3.45, {\it c} = 6.42 \r{A}) under the same pressure.

\begin{table}[tb]
	\captionsetup{justification=raggedright, singlelinecheck=false}
	\caption{Calculated unit cell parameters, magnetic stabilization energies and magnetic moments for C22 and C23 structures of Fe$_2$P, along with experimental data from literature}
\begin{ruledtabular}
	\begin{tabular}{lcccc}
	& \multicolumn{2}{c}{C22} & \multicolumn{2}{c}{C23} \\
	\hline
	& {This work} & Expt.{\footnote{Refs. \cite{carlsson1973determination,scheerlinck1978neutron}.}} & This work & Expt.{\footnote{Ref. \cite{senateur1976etude}.}} \\
	\hline
	\it a (\AA) & 5.81 & 5.87 & 5.61 & 5.78 \\
	\it b (\AA) & -- & -- & 3.57 & 3.57 \\
	\it c (\AA) & 3.43 & 3.46 & 6.51 & 6.64 \\
	\it V (\AA$^3$) & 100.2 & 103.1 & 130.4 & 136.9 \\
	$\Delta E^{\text{NM-FM}}$ (eV/f.u.) & 0.22 & -- & 0.18 & -- \\
	$\mu^\text{Fe1}$ ($\mu_\text{B}$/atom) & 0.82 & 0.96 & 0.18 & -- \\
	$\mu^\text{Fe2}$ ($\mu_\text{B}$/atom) & 2.23 & 2.31 & 1.61 & -- \\
	$\mu^\text{f.u.}$ ($\mu_\text{B}$/f.u.) & 3.01 & 3.27 & 1.75 & -- \\
	\label{table1}
\end{tabular}
\end{ruledtabular}
\end{table}
		
Magnetic stabilization energy ($\Delta E^{\text{NM-FM}} = E^\text{NM} - E^\text{FM}$), evaluated as the difference between total energies of non-magnetic (NM) and ferromagnetic (FM) configurations, is 0.22 eV/f.u. for C22, which is in good agreement with the previous DFT result {\cite{scott2008p}}, and 0.18 eV/f.u. for C23 structure, for which there are no earlier reports for comparison. Thus our calculations establish the stability of FM configuration than NM for both phases of Fe$_2$P. Further, calculated total density of electronic states (Fig. S1, Ref. \cite{suppl}), corroborate the metallic and magnetic characteristics of both phases. Calculated magnetic moments for C22 phase are 0.82 and 2.23 $\mu_\text{B}$ for Fe(I) and Fe(II), respectively. The total magnetic moment per formula unit is 3.01 $\mu_\text{B}$, owing to small moments induced at P sites due to the polarization effect of the Fe atoms, which are antiparallel to the Fe moments. These values are in good agreement with the previous experimental and theoretical results {\cite{scheerlinck1978neutron,tobola1996magnetism,ishida1987electronic,liu2013fe}}. For C23 phase, calculated magnetic moments are 0.18 and 1.61 $\mu_\text{B}$ for Fe(I) and Fe(II), respectively, and the total magnetic moment is ~40$\%$ lower than that for C22 phase. Since experimental or theoretical data are not available for magnetic moments of C23 structure considered in the present work, we cannot make a comparison. However, the average magnetic moment per Fe atom (~0.84 $\mu_\text{B}$) is in reasonable agreement with the earlier DFT result (0.7 $\mu_\text{B}$) \cite{nisar2010equation}.

Pressure dependent studies of materials are essential for the understanding of phase transition mechanism and structural stability. To examine the stability of Fe$_2$P under hydrostatic pressure, the total energy as a function of volume for both the phases of Fe$_2$P is determined and the results are shown in Fig. \ref{fig1}, together with enthalpy of C23 phase with reference to C22 phase as a function of hydrostatic pressure (inset of Fig. \ref{fig1}). Also, variation of magnetic moment with pressure is illustrated in Fig. S2 of Ref. \cite{suppl}. As evident from the Fig. \ref{fig1}, no phase transformation occurs between C22 and C23 phase under compression up to 80 GPa. This is in disagreement with the computational results presented in the Ref. \cite{nisar2010equation}, where they predict C22 to C23 phase transformation under a pressure of 26 GPa. However, our results are in accordance with the experimental observations {\cite{dera2008high,gu2013probing}}, where C22 to C23 structural transformation under pressure is observed only when the samples are heated to higher temperatures of $\sim1400$ K.
\begin{figure}[bt]
	\centering	 
	\includegraphics[scale=0.35]{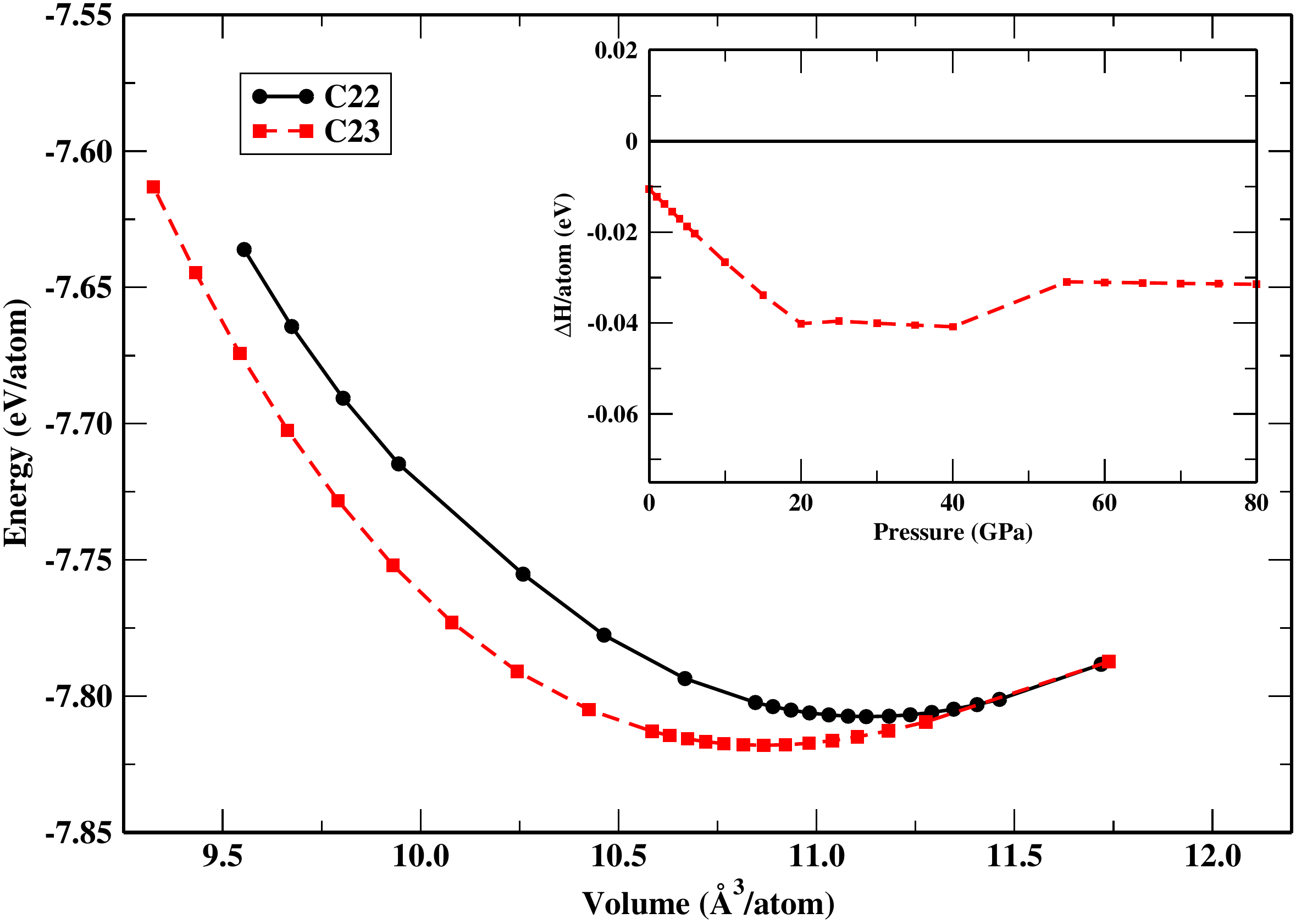}
	\captionsetup{justification=raggedright, singlelinecheck=false}
	\caption{\label{fig1} (color online). Variation of energy as a function of volume for C22 and C23 phases of Fe$_2$P. The inset: calculated enthalpy of C23 phase with reference to C22 phase as a function of pressure.}
\end{figure}

\begin{table*}[t]
	\captionsetup{justification=raggedright, singlelinecheck=false}
	\caption{\label{table2}
		Calculated unit cell parameters for C22 and C23 phases of Fe$_2$P, using GGA, GGA+vdW, GGA+U, LDA and LDA+U along with experimental values for comparison.}
	\begin{ruledtabular}
		\begin{tabular}{lcccccc}
			& Expt.{\footnote{Experimental values are taken from Refs. \cite{carlsson1973determination} and \cite{scheerlinck1978neutron} respectively for C22 and C23 phases.}} & GGA & GGA+vdW & GGA+U & LDA & LDA+U \\
			\hline
			C22 \\
			\hline
			\it a (\AA) & 5.87 & 5.81 & 5.77 & 5.82 & 5.57 & 5.68 \\
			\it b (\AA) & 3.46 & 3.43 & 3.39 & 3.42 & 3.42 & 3.36 \\
			\it V (\AA$^3$) & 103.1 & 100.2 & 97.6 & 100.3 & 91.8 & 93.7 \\
			\hline
			C23 \\
			\hline
			\it a (\AA) & 5.78 & 5.61 & 5.55 & 5.95 & 5.46 & 5.45 \\
			\it b (\AA) & 3.57 & 3.57 & 3.54 & 3.45 & 3.50 & 3.50 \\
			\it c (\AA) & 6.64 & 6.51 & 6.47 & 6.59 & 6.38 & 6.37 \\
			\it V (\AA$^3$) & 136.9 & 130.4 & 126.9 & 135.3 & 121.9 & 121.4 \\
			$\Delta E$ (meV/atom) & -- & 10.4 & 19.6 & 9.7 & 36.8 & 25.8 \\
		\end{tabular}
	\end{ruledtabular}
\end{table*}
As evident from the Fig. \ref{fig1}, C23 phase has lower energy than the C22 phase under the calculation conditions of absolute temperature and pressure, and the energy difference, i.e. $\Delta E_0 = E^\text {C23}_0 - E^\text{C22}_0$ is calculated to be -10.4 meV/atom. To cross check this energy difference and to validate the functional used in our DFT calculations, we performed structural optimization using local density approximation (LDA). Also, calculations are carried out with on-site Hubbard corrections with U = 2 eV and J = 0.95 eV applied to Fe atoms {\cite{wu2010first}} for both GGA (GGA+U) and LDA (LDA+U). Further, weak interlayer van der Waals interaction (GGA+vdW) is accounted using zero damping DFT-D3 method of Grimme {\cite{grimme2010consistent}}. The values are tabulated in Table \ref{table2}, which confirm our results obtained using GGA. Thus our DFT calculations predict C23 phase as the ground state structure of Fe$_2$P, under absolute conditions of temperature and pressure. This is in agreement with the theoretical results of Wu {\it et al.} \cite{wu2010first}, howbeit contradicting the experimental observation. A possible reason for the above discrepancy between experiments and theory could be zero-point vibrational energy contribution.

\subsection{Phonons}
To elucidate the role of zero-point vibrational energy with respect to the observed stability of C22 phase and to investigate local structural stability of the phases, full phonon dispersions for both phases are determined. Phonon dispersion and density of states for both phases are illustrated respectively in Figs. S3 and S4 of Ref. \cite{suppl}. From figures, it is clear that both C22 and C23 phases are stable as these structures display no unstable modes. To examine the relative stability of Fe$_2$P phases at finite temperatures, we determine Helmholtz free energy. On the assumption that the entropy of the electronic part can be neglected, the Helmholtz free energy, $F$ at volume $V$ and temperature $T$ can be approximated as \cite{lodziana2003dynamical}
\begin{equation}
F(V,T) = E_\text{DFT}(V) + F_\text{vib}(V,T),
\end{equation}
where $E_\text{DFT}(V)$ is the total energy evaluated from the electronic structure calculations. $F_\text{vib}(V,T)$ is the vibrational contribution to the free energy given by the equation \cite{togo2015first}
\begin{equation}
F_\text{vib}(V,T) = \frac{1}{2} \sum_{\text{qj}}\hbar \omega_{\text{q}j} + k_BT\sum_{\text{qj}}\ln\left[1-\exp\left(\hbar\omega_{\text{qj}}/k_\text {B}T\right)\right]
\end{equation}

where $T$, $k_{\text{B}}$ and $\hbar$ respectively are temperature, Boltzmann constant and reduced Planck constant. $\omega_\text{qj}$ is phonon frequency of the phonon mode labelled by a set \{q, j\}, where q and j denotes wave vector and band index, respectively.  The first term in the above equation is the zero-point vibration energy (EZPV), which is calculated to be 44.6 meV/atom and 49.2 meV/atom, respectively for C22 and C23 phases. Calculated $F_\text{vib}$ and $F(V,T)$ as a function of temperature for both phases are shown in Fig. \ref{fig2}. It is clear from the figure that C22 phase becomes stable at $\sim$300 K due to thermal vibrations and therefore confirms the stability of C22 phase over C23 phase under zero pressure conditions at room temperature. However, experimentally Fe$_2$P is reported to have C22 structure at temperatures down to 10 K \cite{jernberg1984mossbauer}. This implies that the zero-point vibrational energy difference i.e. $\Delta E_\text{ZPV}$ = $E^\text{C23}_\text{ZPV}$ - $E^\text{C22}_\text{ZPV}$, which is calculated to be 4.6 meV/atom, favors C22 phase, yet not large enough to stabilize the C22 phase.

\begin{figure*}
	\centering
	\begin{subfigure}[t]{.5\textwidth}
	    \centering
		\includegraphics[scale=0.65]{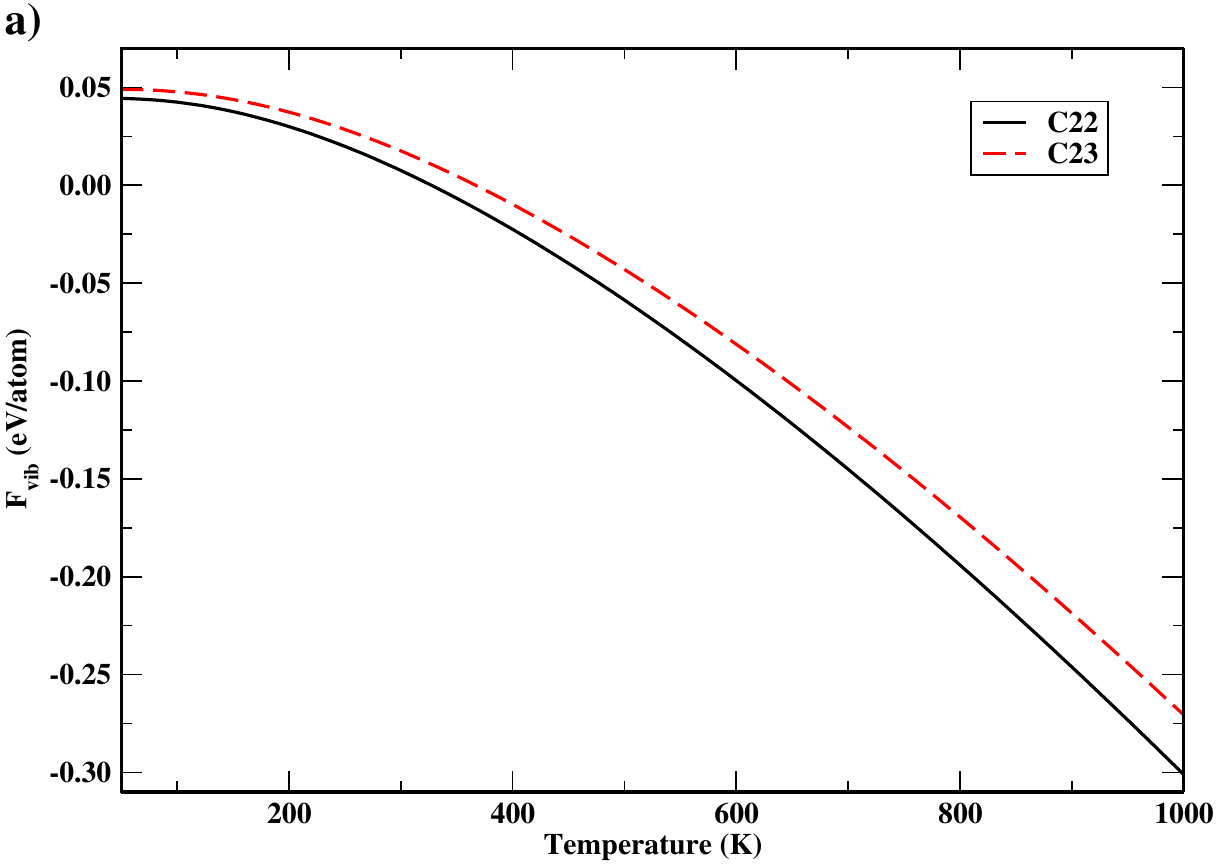}
		\label{fig2a}	
	\end{subfigure}%
	\begin{subfigure}[t]{.5\textwidth}
		\centering
		\includegraphics[scale=.65]{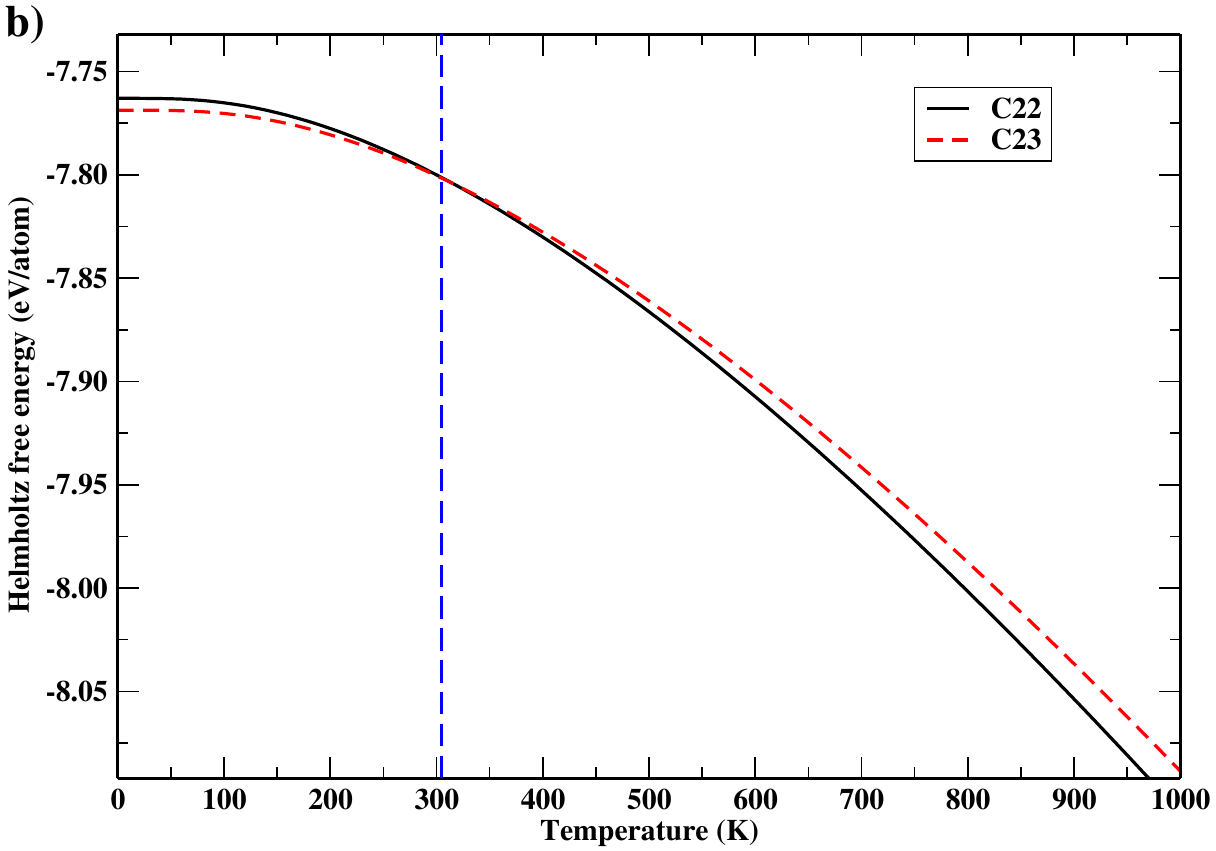}
		\label{fig2b}	
	\end{subfigure}
	\captionsetup{justification=raggedright, singlelinecheck=false}
	\caption{\label{fig2} (color online). Calculated (a) vibrational contribution to the free energy, and (b) Helmholtz free energy for C22 and C23 phases of Fe$_2$P.}
\end{figure*}

\subsection{Spin fluctuations}
\label{sec3c}
From the above arguments, it is clear that electronic and vibrational energies together cannot explain the stability of C22 phase of Fe$_2$P, which is established as the stable phase by experiments at temperatures close to 0 K. Hence we have to seek for a distinct aspect of physics that has not been accounted so far in our calculations, but plays a key role in determining the ground state structure of Fe$_2$P. It is well known that in Fe-pnictide systems spin fluctuations are very important. For example, the pairing mechanism in recently discovered Fe-pnictide based superconductors such as LaFePO, LaFeAsO$_{1-x}$F$_x$ etc. are explained through spin fluctuations {\cite{mcqueen2008intrinsic,singh2008density,kohama2008ferromagnetic,yao2009spin}}. Besides, for Si-doped Fe$_2$P, Delczeg-Czirjak {\it et al.} \cite{delczeg2010ab} reported zero-point spin fluctuation (ZPSF) energy difference to be large enough to stabilize the hexagonal phase over DFT predicted body centred orthorhombic phase. To explore whether C22 phase could possibly be stabilized by ZPSF at very low temperature, we write the total energy of the system as
\begin{equation}
E = E_\text{DFT}+E_\text{ZPV}+E_\text{ZPSF},
\end{equation}
where $E_\text{DFT} $ is the electronic energy obtained from DFT calculation, $E_\text{ZPV}$ is the energy correction due to zero-point vibration and $E_\text{ZPSF}$ is zero-point spin fluctuation energy which is given by the equation {\cite{delczeg2010ab,solontsov1995zero,solontsov2009magnetism}} (refer APPENDIX A)
\begin{equation}
E_\text{ZPSF}=\frac{3\hbar}{4\pi}\omega_\text{SF} ln(1+(\omega_\text c/\omega_\text{SF})^2
\label{zpsfeqn}
\end{equation}
where $\omega_\text c$ is cutoff frequency and $\omega_\text{SF}$ is characteristic frequency of the spin fluctuation. In our study we use the approximations, $\hbar\omega_\text{SF} = \chi^{-1}\mu_\text B^2$ and $\hbar\omega_\text c >> k_\text BT_\text{melt}$ \cite{solontsov2009magnetism}, where $\chi$ and $T_\text{melt}$ are magnetic susceptibility and melting tempertaure, respectively. In order to calculate $E_\text{ZPSF}$ term, we determined total energy as a function of constrained magnetic moment for both phases of Fe$_2$P, as shown in Fig. \ref{fig3}.

\begin{figure}[b]
 \includegraphics[scale=0.35]{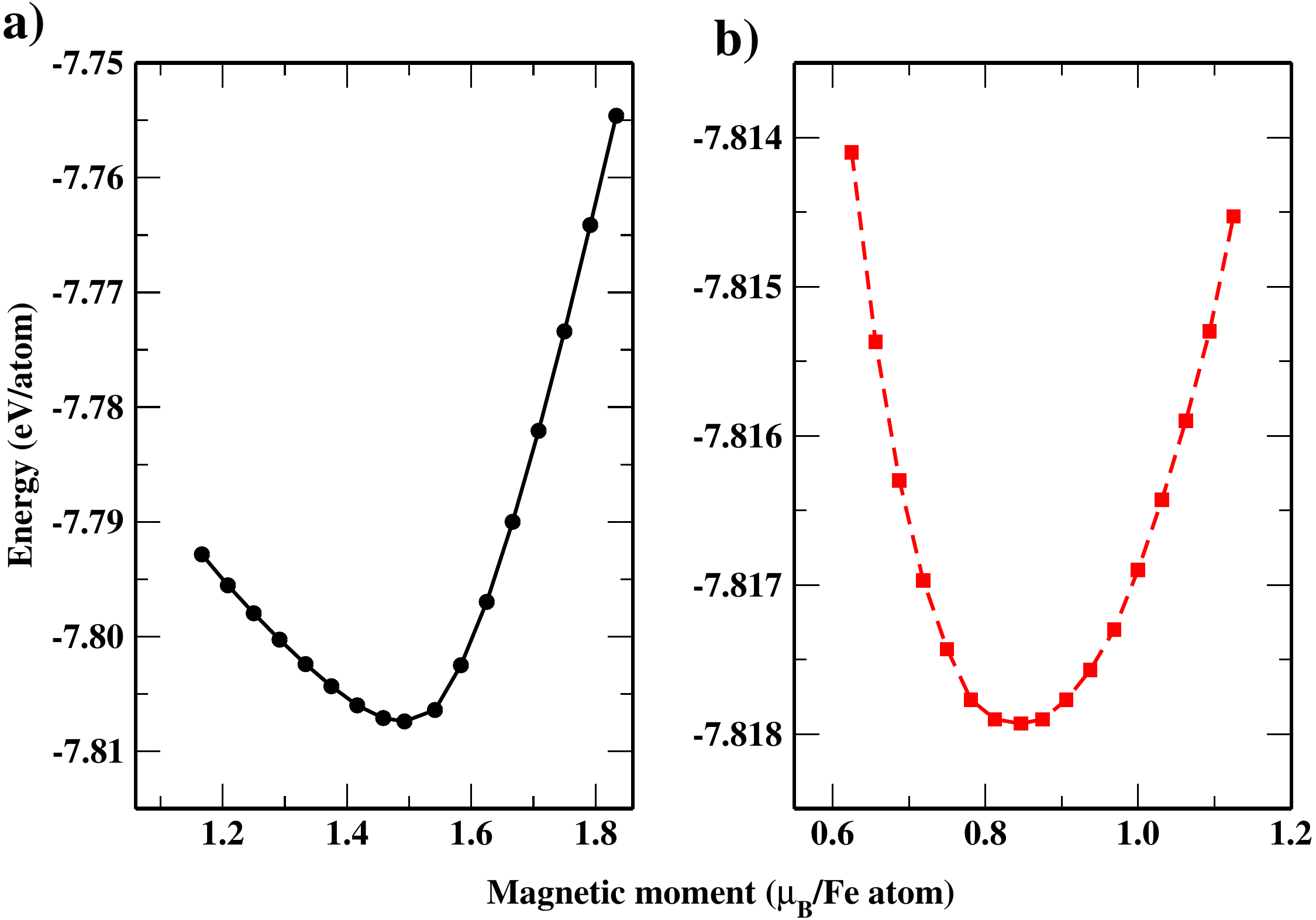}
 \captionsetup{justification=raggedright, singlelinecheck=false}
 \caption{\label{fig3} (color online). Energy as a function of magnetic moment per Fe atom calculated for (a) C22, and (b) C23 phases of Fe$_2$P.}
\end{figure}

The magnetic susceptibility $\chi$ is calculated using the relation \cite{ortenzi2012accounting}
\begin{equation}
\chi^{-1} = \left. \frac {\delta^2E}{\delta M^2} \right|_{M = M_0},
\end{equation}
where $E$ is the total energy evaluated from the electronic structure calculations, $M$ is the magnetic moment and $M_0$ is its value at equilibrium. Considering the lower limit for cutoff frequency, $\hbar\omega_\text c = \hbar\omega_\text c^\text{min} = k_\text BT_\text{melt}$, we obtain ZPSF energy difference i.e. $\Delta E_\text{ZPSF} = E^\text{C23}_\text{ZPSF} - E^\text{C22}_\text{ZPSF}$ equal to 23.6 meV/atom. Accordingly we can conclude that the ZPSF energy difference is large enough to overcome the DFT predicted energy difference (i.e. -10.4 meV/atom) and thus stabilizes the C22 phase. For the precise value of ZPSF energy, however, more accurate values of characteristic frequencies and cut-off frequencies are required. Typically, zero-point effects are crucial in the context of stability of layered or 2D materials such as boron nitride {\cite{albe1997theoretical,furthmuller1994ab}}. However, our calculations disclose a rare case where quantum effects play pivotal role in the structural stability of a 3D material. Based on above arguments, we may predict the possibility of magnetic field induced structural quantum phase transition in Fe$_2$P. This work may provide a guide to further theoretical and experimental investigations on quantum phase transition in Fe$_2$P.
\section{Conclusions}
In summary, we have systematically examined the ground state structural stability of Fe$_2$P with respect C22 and C23 phases, using first-principles calculations based on DFT. Calculated lattice parameters and magnetic properties for both phases agree well with experimental and theoretical results reported earlier. No phase transition between C22 and C23 phases is noticed under pure hydrostatic pressure, at 0 K temperature, in accordance with the experimental findings. Orthorhombic C23 phase is found to have lower energy than C22 phase, negating the experimental observation. Estimated zero-point vibrational energy is predicted to favor the C22 phase, yet insufficient to stabilize the hexagonal phase. The observed ground state stability of C22 phase is then attributed to lower ZPSF energy of C22 phase relative to C23. Thus, we present a unique result where quantum effects are found to significantly influence the structural stability of a bulk material. Experimental verification and deep theoretical understanding of the possible quantum phase transition can be envisioned as the future scope of this work.
\section*{Acknowledgement}
We acknowledge support from the Convergence Agenda Program (CAP) of the Korea Research Council of Fundamental Science and Technology (KRCF) and Global Knowledge Platform (GKP) program of the Ministry of Science, ICT and Future Planning.
\section*{Appendix A: Derivation of the zero point spin fluctuation energy term ($E_\text{ZPSF}$)}
\renewcommand{\theequation}{A.\arabic{equation}}
Suppose $\mathbf{m}$ is the total magnetic moment of the unit cell and let us consider that $\mathbf{m}$ fluctuates around its mean value $\langle \textbf{m}\rangle$. Therefore we can write,
\begin{equation}
\mathbf{m}=\langle \textbf{m}\rangle+\mathbf{\delta m}
\end{equation}
The magnetic energy of the system can be expressed in terms of Ginzburg-Landau expansion,
\begin{equation}
\begin{split}
E_{\mathbf{m}}=& a_0+\frac{1}{2\chi}\mathbf{m}^2+\frac{a_2}{4}\mathbf{m}^4+...\\
=&a_0+\frac{\langle\textbf{m}\rangle^2}{2\chi}+\frac{a_2\langle \textbf{m}\rangle^4}{2}+\frac{\mathbf{\delta m}^2}{2\chi}+\\
&\frac{a_2 \mathbf{\delta m}^4}{2}+\frac{\mathbf{\delta m}\langle\textbf{m}\rangle}{\chi}  +...  
\end{split} 
\end{equation}
where $\chi$ is the spin susceptibility of the system, which can be calculated using fixed moment calculations in the framework of the density functional theory \cite{goh2017publisher} as shown in Sec. \ref{sec3c}, and a$_0$ is the non-magnetic contribution to the energy. Now setting terms which are odd such as $\frac{\mathbf{\delta m}\langle\textbf{m}\rangle}{\chi}$ to zero (as they would break the time reversal symmetry) as well as neglecting the higher order terms, we get,
\begin{align}
E_{\mathbf{m}}=&a_0+\frac{\langle\textbf{m}\rangle^2}{2\chi}+\frac{a_2\langle \textbf{m}\rangle^4}{2}+\frac{\mathbf{\delta m}^2}{2\chi}+\frac{a_2 \mathbf{\delta m}^4}{2}\nonumber \\
& = E_{\langle\textbf{m}\rangle}+E_\text{SF}
\end{align}
In the limit of small fluctuation ($\mathbf{\delta m} \longrightarrow 0)$ the spin fluctuation energy $E_\text{SF}$ is given by,
\begin{equation}
E_\text{SF} \simeq \frac{1}{2\chi}\mathbf{\delta m}^2
\end{equation}
The fluctuation can be split in to two parts,
\begin{equation}
\mathbf{\delta m}^2=\mathbf{\delta m}^2_\text{ZP}+\mathbf{\delta m}^2_\text{Th}
\label{flu}
\end{equation}
The first term in the Eq. \ref{flu} is the zero temperature fluctuation, while the second term is due to fluctuation at finite temperature. Since we are interested only about the fluctuations close to 0 K, we can write,
\begin{equation}
E_\text{ZPSF}\simeq \frac{1}{2\chi}\mathbf{\delta m}_\text{ZP}^2
\label{Energy}
\end{equation}
The remaining task is to calculate $\mathbf{\delta m}_\text{ZP}^2$, which can be related to dynamic spin susceptibility $\chi(q,\omega)$ via the fluctuation dissipation theorem given by,
\begin{equation}
\mathbf{\delta m}^2_\text{ZP}=3\hbar\int dq \int \frac{d\omega}{2\pi}\frac{1}{2}Im\chi(q,\omega)
\label{FDT}
\end{equation}
Now let us assume that all the fluctuations are localized near Fe-sites and therefore the spin susceptibility due to such fluctuations would be dispersion-less, i,e $\chi(q,\omega)\longrightarrow \chi(\omega)$. The form of the dynamic spin susceptibility can be obtained from the Ref. \cite{solontsov2005spin} is given by,
\begin{equation}
\chi(\omega)=\frac{\chi}{1-i\omega/\omega_\text{SF}}
\end{equation}
where $\chi$ is the frequency independent Kohn-Sham spin susceptibility, obtained from the fixed moment calculation. The Eq. \ref{FDT} now becomes,
\begin{align}
\mathbf{\delta m}^2_\text{ZP}=&\frac{3\hbar\chi}{4\pi }\int^{\omega_\text c}_{0} d\omega \frac{\omega/\omega_\text{SF}}{1+(\omega/\omega_\text{SF})^2} \nonumber\\
=&\frac{3\hbar\chi}{2\pi}\omega_\text{SF} ln(1+(\omega_\text c/\omega_\text{SF})^2)
\end{align}
where $\omega_\text{SF}$ is the characteristic frequency of the spin fluctuation and the energy integration is performed upto a cut-off frequency of $\omega_\text c$. The spin-fluctuation energy is therefore given by from the Eq. \ref{Energy},
\begin{equation}
E_\text{ZPSF}=\frac{3\hbar}{4\pi}\omega_\text{SF} ln(1+(\omega_\text c/\omega_\text{SF})^2)
\label{Final}
\end{equation}
The Eq. \ref{Final} was used in the manuscript in order to estimate the contribution of spin fluctuation. We have not considered any  anharmonic effects as we are interested in the temperature region far away from the magnetic transition temperature. It was pointed out in the previous study \cite{solontsov2005spin} that spin anharmonicity is important only near the magnetic transition temperature.
\bibliography{Fe2P_Bibtex}

\end{document}